\documentclass[aps,preprint]{revtex4}%
\usepackage{amsfonts}
\usepackage{amsmath}
\usepackage{amssymb}
\usepackage{graphicx}%

\usepackage[T1]{fontenc}

\begin{document}
\preprint{ }
\title[SFA without tunneling]{Electrodynamics of the strong-field approximation and deficiencies of the
tunneling model}
\author{H. R. Reiss}
\affiliation{Max Born Institute, Berlin, Germany}
\affiliation{American University, Washington, DC, USA}

\pacs{PACS number}

\begin{abstract}
The strong-field laser physics enterprise is investing important resources in
the study of the effects of oscillatory electric fields on matter using the
tunneling concept, whereas laser fields are vector fields that do not support
the tunneling model. Oscillatory electric fields and propagating plane-wave
laser fields are different electrodynamic phenomena, and similarities in their
effects diminish as field intensity increases. Major differences are known in
the case of very low frequencies where oscillatory electric fields act
adiabatically as the frequency declines, in contrast to extreme relativistic
effects of strong laser fields. Many supposed new phenomena, such as ATI
(Above-Threshold Ionization), channel closing, and stabilization were studied
in terms of propagating fields before they came to the attention of the atomic
physics community. This illustrates the efficiency of using proper
electrodynamic methods for the treatment of laser effects. Problems arising
from the conflation of the effects of oscillatory electric fields and
propagating fields are exacerbated by using ill-defined nomenclature, such as
KFR and SFA, that has been used indiscriminately to apply both to oscillatory
electric fields and to propagating fields. Research resources can be applied
with much-improved efficiency if proper electrodynamic treatment of laser
fields is employed.

\end{abstract}
\date[26 August 2019]{}
\maketitle

\section{Introduction}

Laser fields are transverse fields that propagate in vacuum at the speed of
light. The widely-used dipole approximation (DA) means that the speed of
propagation is zero; the laser field is being modeled as an oscillatory
electric field. This replacement works well when the laser field is only a
small perturbation, but when the laser is sufficiently strong that it is the
dominant influence on a photoelectron, the need to treat properly its
electrodynamic character becomes of increasing importance.

Many strong-laser researchers know little of the history of strong-field
physics prior to the historic Saclay experiment \cite{ati} that detected the
above-threshold ionization (ATI) phenomenon. The origins of strong-field
physics are to be found in work that was done \cite{hrdiss,hr62} about 20
years before the Saclay experiment. Earlier work by Sengupta \cite{sengupta}
was not known to the Princeton group of John Wheeler and John Toll, when they
suggested the use of the Volkov solution \cite{volkov} as a tool to
investigate the reasons why the new covariant perturbation theory of Feynman
and Dyson is not convergent \cite{dyson}.

That early work is entirely relativistic, treating properly the plane-wave
field as a propagating field. The ATI effect is inherent in the formalism. The
investigation of the convergence properties of external-field electrodynamics
demonstrated an upper limit to convergence set by the first channel closing
\cite{hrdiss,hr62}. \textquotedblleft Channel closing\textquotedblright%
\ refers to the situation when the ponderomotive energy that must be provided
in a strong-field process becomes sufficiently large that the minimal number
of photons needed for a process indexes upward. Channel closing (or low-order
peak suppression) in atomic ionization was later investigated in Refs.
\cite{hr80} and \cite{mullertipvdw}.

Atomic stabilization by strong fields was remarked as long ago as 1970
\cite{hr70}.

This early identification of explicit strong-field phenomena is testimony to
the advantages of treating the laser field in appropriate terms.

There have been two fundamentally different approaches to the description of
atomic ionization by strong laser fields. One approach builds on the success
achieved in perturbative AMO (Atomic, Molecular, Optical) physics using the DA
(Dipole Approximation). This approximation neglects the magnetic component of
a laser field, thus treating the laser field as if it were an oscillatory
electric field. When the laser field is only a perturbation as compared to the
binding potential that attaches an atomic electron to the parent atom, this
method has been very successful. When the extension of this approach to strong
laser fields is done, the analytical approximation thus formed is designated
herein as an SEFA (Strong Electric Field Approximation). When the laser field
is nonperturbatively strong, the laser becomes the dominant influence. An
analytical approximation that treats the laser field as a transverse (or
propagating) field as the dominant influence on the photoelectron is here
designated as an SPFA (Strong Propagating Field Approximation). This article
examines the relative merits of these two approaches from the point of view of
basic electrodynamics. This goal carries with it the need to distinguish
between SEFA and SPFA approaches, since long-term misdirection on this matter
is shown by examples to have led to costly delays in understanding
strong-field effects.

The following two Sections contain important preliminaries. First, the Maxwell
equations appropriate to propagating fields and to oscillatory electric fields
are contrasted. The low frequency limit of the laser field is where the SEFA
fails completely for the description of laser effects. Then the problem of
nomenclature is discussed, since this has been the source of serious
misunderstandings. The main body of the article follows these preliminaries:
comparisons between SEFA and SPFA\ applicabilities and constraints. Evidence
of SEFA-caused major delays in the development of strong-field physics is summarized.

The descriptions \textit{transverse field}, \textit{propagating field},
\textit{plane-wave field} and \textit{laser field }are used interchangeably here.

All electromagnetic quantities are expressed in Gaussian units.

\section{Preliminaries}

The Maxwell equations governing the behavior of laser fields are so different
from the equation governing an oscillatory electric field that it is evident
that these two manifestations of electrodynamic behavior can become very
different outside a limited range of correspondence. These differences are
most obvious as the frequency declines, so that contrast will be made in this Section.

\subsection{Maxwell equations}

The vacuum Maxwell equations for the electric field $\mathbf{E}$ and the
magnetic field $\mathbf{B}$ of a propagating field have no source terms, so
they are simply%

\begin{align}
\mathbf{\nabla\cdot E}  &  =0,\label{A}\\
\mathbf{\nabla\cdot B}  &  =0,\label{B}\\
\mathbf{\nabla\times B}-\frac{1}{c}\partial_{t}\mathbf{E}  &  =0,\label{C}\\
\mathbf{\nabla\times E}+\frac{1}{c}\partial_{t}\mathbf{B}  &  =0. \label{D}%
\end{align}
The unique solution for these source-free equations is a propagating,
transverse field.

For a laser field described within the DA, the constraints that the electric
field has no spatial dependence and the magnetic field is not present, reduce
the four completely-stated Maxwell equations to the single equation
\cite{hrlg}%
\begin{equation}
\partial_{t}\mathbf{E}^{DA}\left(  t\right)  =-4\pi\mathbf{J}^{DA}\left(
t\right)  , \label{E}%
\end{equation}
where $\mathbf{J}^{DA}\left(  t\right)  $ is a virtual source current that
must be postulated to have a nontrivial result.

The electric field $\mathbf{E}^{DA}$ is a proxy field that, over a limited
range of parameters, mimics the effect of a propagating field. The structure
of four source-free Maxwell equations is so different from the single
source-dependent Maxwell equation for the proxy field that mimicry is never
exact, and can be quite limited in extent.

When authors who employ the SEFA refer to a photoelectron being
\textquotedblleft driven\textquotedblright\ back to the remnant ion after
photoionization, the \textquotedblleft driving\textquotedblright\ comes from
the virtual source current $\mathbf{J}^{DA}\left(  t\right)  $. There are no
sources for a photoelectron in a laser field. A charged particle in a
propagating field executes a periodic motion (see Ref. \cite{ss} for a full
description), but this motion simply repeats itself for each period of the
plane wave. There is no net transfer of energy and momentum after a full
period, such as can follow from the DA source term $\mathbf{J}^{DA}\left(
t\right)  $. For the first full period of the laser field after ionization by
a linearly polarized laser occurs, the motion described within the SEFA is
nearly identical to the free particle motion in a plane wave, which explains
the quantitative success of the rescattering model of the SEFA
\cite{kuchiev,kulander,corkum}. However, motion beyond the first period will
entail increasing and unexplored discrepancies between SEFA and SPFA.

\subsection{Low frequency limit}

\subsubsection{SEFA}

The SEFA is quite simple as $\omega\rightarrow0$. An oscillatory electric
field approaches a constant electric field in the zero-frequency limit. The
property of \textit{adiabaticity} refers to slowing of changes in the system
in this limit. This was first discussed by Keldysh \cite{keldysh}, and the
parameter he defined to measure approach to this limit is called the
adiabaticity parameter or, equivalently, the Keldysh parameter $\gamma_{K}$,
defined as%
\begin{equation}
\gamma_{K}=\sqrt{E_{B}/2U_{p}}, \label{F}%
\end{equation}
where $E_{B}$ is the initial binding energy of a photoelectron in the parent
atom. An alternative name for $E_{B}$ is ionization potential $IP$ or $I_{p}$.
This behavior is said to be well-confirmed by alternative schemes for
calculating low-frequency electric-field phenomena, and by numerical solution
of the time-dependent Schr\"{o}dinger equation, conventionally referred to as
TDSE. TDSE calculations are widely regarded as exact, thus supplying a
putative definitive test of analytical approximations.

The problem with all of this widely accepted work is that laser fields do not
behave in that way at all as $\omega\rightarrow0$. TDSE is a DA procedure, and
verifies only oscillatory electric field phenomena, not laser-induced phenomena.

Qualitatively, SEFA behavior as $\omega\rightarrow0$ is a constant electric
field. This stands in contrast to SPFA behavior, where the field propagates at
the speed of light for all frequencies, no matter how low.

\subsubsection{SPFA}

As shown in Ref. \cite{hr101,hrtun}, declining frequency does not lead to
adiabatic behavior. When the field is intense, it leads first to a domain
where the magnetic component of the laser field becomes significant, with a
further decline in frequency leading to fully relativistic behavior. It is
this contrast in the properties of oscillatory electric fields in comparison
with fully-described plane-wave fields that demonstrates the complete failure
of the proxy field in Eq. (\ref{E}) to mimic the laser field as described by
Eqs. (\ref{A}) - (\ref{D}).

The failure of the DA at low frequencies has far more practical significance
than the well-known failure of the DA at high frequencies. The onset of this
low-frequency failure can be detected even at a wavelength of $1.51\mu m$, as
will be shown below.

An example of a plane wave phenomenon of extremely low frequency is the
Schumann resonance \cite{schumann}. This is a naturally occurring phenomenon
in which powerful lightning strikes generate low frequency radio waves that
resonate in the cavity formed by the Earth's surface and the ionosphere. The
lowest mode of this cavity is $7.83Hz$, corresponding to a wavelength about
equal to the circumference of the Earth. On a laboratory scale, a plane wave
with a wavelength equal to the circumference of the Earth would appear to be a
constant field. Nevertheless, a constant electric field cannot spread its
influence over the entire planet, thus emphasizing the fundamental difference
between electric fields and propagating fields.

\section{Nomenclature}

The naming of objects or procedures must be precise in all the sciences, so
that the name employed has a universally understood meaning. Unfortunately,
observance of this basic requirement has been violated for many years in
strong-field physics. This matter is of quintessential importance for the
study of strong-field laser phenomena, since the distinctions between SPFA and
SEFA behavior have remained almost unseen and unaddressed for 40 years.

\subsection{KFR}

KFR refers to the names of three authors (Keldysh \cite{keldysh}, Faisal
\cite{faisal}, and Reiss \cite{hr80}). This conflates a low-frequency SEFA
method \cite{keldysh}, a high-frequency SEFA method \cite{faisal}, and an SPFA
method \cite{hr80}. The term \textquotedblleft KFR\textquotedblright\ was
introduced in Ref. \cite{bucks2} to refer generically to nonperturbative
analytical approximations. The continued use of the KFR acronym is unfortunate
in view of the inequivalent approximations that gave rise to the terminology.

\subsubsection{The Keldysh approximation}

The essence of the Keldysh approximation is that everything is treated within
the DA; the interaction Hamiltonian is stated in the length gauge; the method
used is equivalent to a time-reversed S-matrix theory; and the final state
$\Psi_{f}$ is replaced by a Volkov state with a gauge-transformation factor
introduced to transform the Volkov solution expressed in the Coulomb gauge to
the length gauge instead.

The limitations imposed by the approximations made by Keldysh as applied to
laser fields can be stated concisely. The DA fails at both high and low field
frequencies. The $\mathbf{r\cdot E}$ potential of the length gauge restricts
the field to being purely an oscillatory electric field \cite{hrlg}. The
\textquotedblleft dipole-approximated Volkov solution\textquotedblright\ is
not a Volkov solution in the usual sense of describing the behavior of a free
electron in a propagating field. The Keldysh approximation is an approximation
for ionization or detachment of an electron from a bound state by a strong,
low (but not very low) frequency, oscillatory electric field.

Keldysh, understandably unaware in 1964 of the properties of the S-matrix,
inappropriately regards the non-interacting nature of the initial state
$\Phi_{i}$ as an additional approximation.

The Keldysh approach was recast in Refs. \cite{nikrit1,ppt,adk} as ionization
by a low-frequency oscillatory electric field using an adiabatic
approximation, rather than with Volkov-solution concepts.

\subsubsection{The Faisal approximation}

In the Faisal approximation \cite{faisal}: Everything is treated within the
DA, the interaction Hamiltonian is expressed in the velocity gauge, and the
direct-time S-matrix transition amplitude is employed.

The difficult problem of finding a suitable approximation for the initial
state $\Psi_{i}$ in the direct-time S-matrix formalism is approached by using
the KH (Kramers-Henneberger) \cite{kramers,henneberger} transformation to a
moving frame of reference, and then replacing that oscillatory orbit with the
assumption that spatial coordinates remain centered at the atomic nucleus.
This is a high-frequency approximation.

The Faisal approximation can be summarized as an approximation for ionization
or photodetachment of an electron from a bound state by a high-frequency
oscillatory electric field.

\subsubsection{The Reiss approximation}

The genesis of the R method \cite{hr80} is to be found in studies with the
relativistic Volkov solution \cite{hrdiss,hr62,hr62b,jehr,hrje}. The intended
application to nonrelativistic problems is approached by using the full Volkov
solution (or Gordon solution \cite{gordon}) in a nonrelativistic,
long-wavelength limit. The Coulomb-gauge interaction Hamiltonian is also
employed in its nonrelativistic form. This nonrelativistic long-wavelength
version of the Coulomb gauge is similar in appearance to the velocity-gauge
expression of the interaction Hamiltonian in the DA; the essential difference
being in the $\mathbf{A}^{2}$ term. In the DA, $\mathbf{A}^{2}\left(
t\right)  $ plays no dynamical role; whereas $\mathbf{A}^{2}\left(  t\right)
$ has a central role to play in the nonrelativistic R theory. (See Refs.
\cite{hrup,hrje}.) This distinction is examined in two 1990 papers. In the
first of these papers \cite{hr90}, the 1980 R formalism is shown to be the
nonrelativistic limit of a Klein-Gordon (i.e. spinless relativistic) treatment
of atomic ionization. The second 1990 paper \cite{hrrel} is a completely
Dirac-relativistic version of the SPFA, where initial- and final-state wave
functions as well as the interaction term are Dirac-relativistic. The
nonrelativistic, long-wavelength limit of the final outcome of the
Dirac-relativistic atomic ionization calculation reduces exactly to the 1980 R
paper \cite{hr80}.

The R approximation is an approximation for ionization or detachment of an
electron from a bound state by a strong, propagating field; i.e. by the field
produced by a laser.

The R approximation is the same as the nonrelativistic SPFA.

Frequency limitations in the SPFA are much milder than with other methods.
This stands in contrast to the low-frequency and high-frequency restrictions
inherent in tunneling methods and the high-frequency-only domain of the
KH-based approximations. An interesting development is that the SPFA produces
results at high frequencies equivalent to the KH-based methods of Faisal and
Gavrila, and to TDSE \cite{bondar}. That is, the SPFA, fundamentally different
from DA methods at low frequencies, becomes equivalent to DA theories at high
frequencies (albeit intensity-limited to nonrelativistic conditions).

\subsection{SFA}

The terminology \textquotedblleft Strong-Field Approximation\textquotedblright%
\ (SFA) was introduced in Ref. \cite{hr90} to refer exclusively to what is
here labeled the SPFA in the belief that this article had made clear the
distinction between SEFA and SPFA methods. This belief was negated by future
articles (see Ref. \cite{lewen1}) that viewed SFA as equivalent to KFR. This
confusion is what necessitates the use of the four-character acronyms SPFA and
SEFA rather than the simpler three-character SFA.

\section{Properties of the SPFA}

It would seem plausible that an investigation of laser-induced effects would
be more efficient with a method based on the propagating field that is the
character of lasers, than with a proxy field that has a limited range of
accurate mimicry. It is also plausible that physical insights gained from a
study of laser effects based on the properties of plane-wave fields would be
more fruitful and reliable than insights achieved with oscillatory electric
fields. These expectations are corroborated by the examples of this Section.

The first laboratory observation of the ATI phenomenon will be referred to as
the "Saclay experiment" \cite{ati}, referring to the laboratory where the work
was done.

\subsection{Prequels to the Saclay experiment}

A recent publication by well-known authorities in strong-field phenomena
begins with the statement \cite{cbc}: \textquotedblleft Early multiphoton
ionization experiments using intense infrared pulses found the then-amazing
result that an ionizing electron often absorbed substantially more photons
than the minimum needed for ionization.\textquotedblright\ This
\textquotedblleft amazement\textquotedblright\ was experienced by atomic
physicists and physical chemists who habitually employed the dipole
approximation. This stands in contrast to the pre-existing corpus of work with
strong propagating fields. ATI is a feature to be expected in strong-field
experiments. Analogs of ATI had been observed in nonperturbative bound-bound
transitions in atoms leading to the statement \cite{hr70}: \textquotedblleft%
...as the intensity gets very high ... higher order processes become
increasingly important.\textquotedblright\ For photon-multiphoton pair
production \cite{hr71}: \textquotedblleft\ ... an extremely high-order process
can ... dominate the lowest order ...\textquotedblright. For interband
transitions in band-gap solids \cite{hjhr}: \textquotedblleft...high-order
processes can be more probable than lower-order processes when the intensity
is sufficiently high.\textquotedblright

The 1980 strong-field paper \cite{hr80}, written before the publication of the
Saclay experiment, described ATI in detail. This paper examined such matters
as the differing shapes of spectra resulting from linear and circularly
polarized lasers, that was not observed until 1986 \cite{bucks}. There was a
detailed description of the channel-closing phenomenon, rediscovered later by
other theoreticians \cite{mullertipvdw}.

There was even a discussion in Ref. \cite{hr80} of the radius of convergence
of perturbation theory, an important topic largely ignored in the strong-field community.

The point of the above remarks is that insights and concrete results
attainable from the point of view of actual laser fields enabled the
prediction and description of explicit strong-laser phenomena that were
unexpected and counter-intuitive within the AMO community.

To the present, it is remarkable that many members of the strong-field
community remain unaware that ATI was predicted in advance of observation, and
that SPFA and SEFA approaches are fundamentally different. This appears to be
a consequence of the KFR and SFA nomenclature problem discussed above, that is
so uncharacteristic of scientific protocols.

\subsection{Quantitative predictions of the SPFA}

The quotation \cite{cbc} cited above about how amazing the ATI phenomenon
appeared to the atomic physics community in 1979, continues with the estimate
that it took until about 1993 before understanding of it was acquired within
the DA. These dates are significant in that, when propagating-field properties
are employed, the qualitative existence of ATI was well-known before 1979.
Essential qualities of ATI became very clear in a 1986 experiment
\cite{bucks}, matched correctly by the SPFA \cite{hr87,hrjosa87}; and
precision experimental results published in 1993 \cite{moh} were precisely
duplicated by the SPFA \cite{hrmoh}.

\subsubsection{Early success}

\begin{figure}
[ptb]
\begin{center}
\includegraphics[
trim=0.000000in 1.504715in 2.132721in 2.134927in,
height=3.1107in,
width=6.9946in
]%
{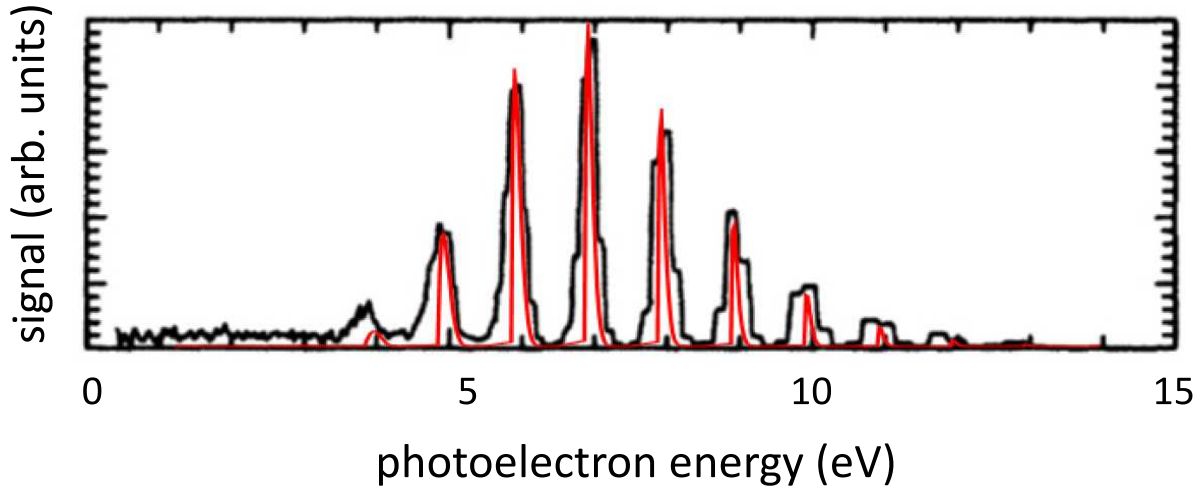}%
\caption{This figure shows the ability of the transverse-field theory of Ref.
\cite{hr80} to replicate the experimental results presented in Ref.
\cite{bucks}. In DA terminology, this is in the multiphoton domain. The black
curve (with wide peaks) is the measured photoelectron spectrum and the red
curve (with narrow peaks) is the theoretical fit. Laser parameters: $1064$
$nm$, peak intensity $2\times10^{13}$ $W/cm^{2}$, pulse duration $100$ $ps$ on
a xenon target ($\gamma_{K}=1.69$). The calculation includes focal averaging
in a Gaussian beam with Gaussian time distribution, and with partial
ponderomotive energy ($U_{p}$) recovery in the very long pulse. The only
fitting parameter employed was the relative fraction of recovered $U_{p}$
(about $80\%$) selected to fix the absolute energy locations of the peaks.}%
\label{fig1}%
\end{center}
\end{figure}

The first numerical correspondence between theory and experiment in
strong-laser problems was from the analysis of a 1986 experiment \cite{bucks}
published in 1987 \cite{hr87,hrjosa87}. The comparison of theory and
experiment is shown in Fig. \ref{fig1}, which is a refinement of the bar-graph
figures of Refs. \cite{hr87,hrjosa87}, and includes focal averaging. The year
1987 was a time at which the debate was ongoing about whether ATI was
something entirely novel or was no more than higher-order perturbation theory
\cite{chinlambr}. This early success of the SPFA attracted only temporary
attention.%

\subsubsection{Precision at high intensity}

\begin{figure}
[ptb]
\begin{center}
\includegraphics[
height=4.395in,
width=3.9738in
]%
{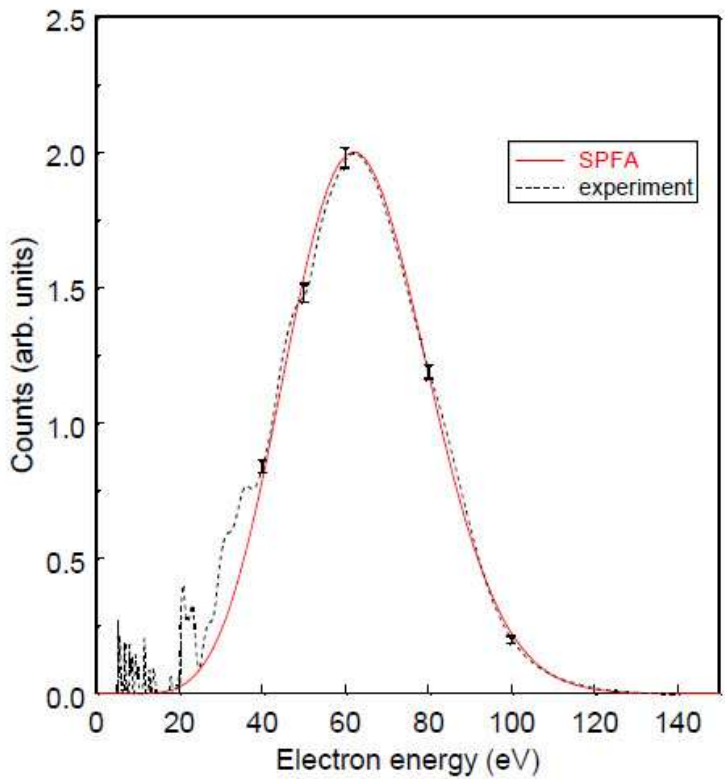}%
\caption{This figure, based on Ref. \cite{hrmoh}, shows the precision possible
with the SPFA of 1980 \cite{hr80} applied to the description of an experiment
\cite{moh} on the ionization of helium at the relatively high intensity of
$1.27\times10^{15}$ $W/cm^{2}$ at $815$ $nm$ wavelength and a pulse length of
$180$ $fs$. By SEFA terminology, this is in the tunneling domain ($\gamma
_{K}=0.40$). The SPFA fit, obtained with a multiphoton theory, is within the
very small experimental error bars. Irregularities in the low energy part of
the spectrum are experimental artifacts \cite{mohpriv}.}%
\label{fig2}%
\end{center}
\end{figure}

Laser capabilities had improved considerably by 1993, with much higher
intensities and shorter pulses attainable. Mohideen, \textit{et al.}
\cite{moh} carried out experiments in which they were able to measure spectra
with impressive accuracy, especially with circular polarization. The only
theoretical predictions available were from the SPFA of Ref. \cite{hr80}, but
Mohideen, \textit{et al.} applied this theory without averaging over the
spatial and temporal intensity distribution that existed in the laser focus.
The same theory was applied again in 1996 \cite{hrmoh}, with focal averaging
incorporated. The results obtained are shown in Fig. \ref{fig2}.%

A conclusion that follows from Fig. \ref{fig1} and Fig. \ref{fig2} is that the
division of frequencies into a tunneling domain and a multiphoton domain has
no meaning with respect to real laser fields. Both of the figures follow from
the same formalism \cite{hr80}, and that formalism involves a sum over
discrete numbers of photons. Figure \ref{fig2} shows the precision possible
with the SPFA when the intensity is high. The curve in Fig. \ref{fig2} appears
to be smooth, with no indication that there are more than 50 photon orders
contained within the single peak.

\subsubsection{Onset of low-frequency failure of the DA}

\begin{figure}
[ptb]
\begin{center}
\includegraphics[
height=3.6512in,
width=5.1102in
]%
{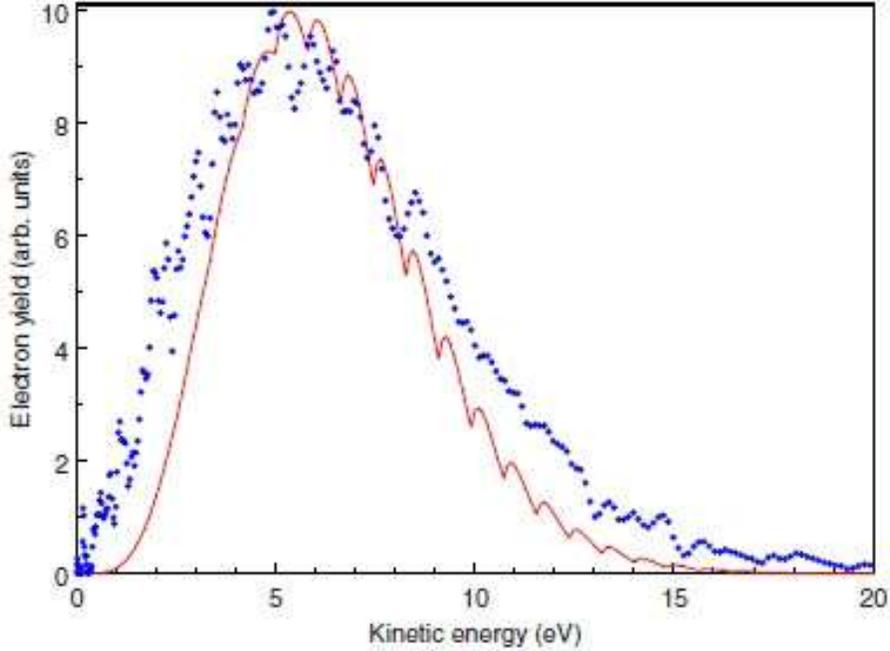}%
\caption{Experimental data \cite{bergues} are shown as dots and the SPFA
calculation \cite{hrberg} as the continuous curve, for the spectrum of
electrons from the photodetachment of the negative fluorine ion ($F^{-}$) by a
laser operating at $1.510$ $\mu m$. The peak laser intensity in the
calculation is taken to be the same as that measured in the laboratory. The
agreement can be regarded as satisfactory, since residual ATI effects are
evident in both the theory and the experiment, and the intensity is not high
enough to expect the precise agreement shown in Fig. \ref{fig2}. TDSE and SEFA
analytical approximations could fit the data only if there was an assumed
$45\%$ increase in peak laser intensity from that found in the laboratory. The
confluence of TDSE and length-gauge analytical calculations occurs because
both use the DA. The assumption of faulty peak intensity measurement is not
sustainable because the measured momentum distribution supports the originally
reported peak intensity. It is the DA that has to be the cause of the conflict
in theoretical approaches \cite{hrberg}.}%
\label{fig3}%
\end{center}
\end{figure}

It has already been noted that major qualitative and quantitative
discrepancies between the SEFA and SPFA develop as the frequency declines. An
important question to be answered is to find the conditions under which those
discrepancies become significant. An initial step in this quest is found in an
experiment \cite{bergues} at Freiburg ostensibly directed towards deciding on
a choice between length-gauge and velocity-gauge analytical approximations.
What was actually being compared was an SEFA and an SPFA analytical
approximation. (Apologies must be made for the adoption of the gauge-choice
misnomer in the article on this matter \cite{hrberg}.) The experiment was
executed with special attention to accurate measurement of the peak laser
intensity that was determined to within a $15\%$ accuracy. Calculations done
with a length-gauge analytical approximation and with TDSE agreed with each
other, but both required a $45\%$ increase over the measured intensity to
achieve a fit to the data. A calculation done with the SPFA of Ref.
\cite{hr80} agreed with the experiment without any alteration of the measured
intensity, but was judged by the Freiburg group to be in error because of the
correspondence between the analytical SEFA and the TDSE. The length gauge was
then proclaimed to be superior to the velocity gauge. However, it was pointed
out \cite{hrberg} that the experimental paper reported a momentum distribution
that supplied an independent measurement of peak intensity, and that agreed
with the initial measurement of the peak intensity. Figure \ref{fig3} shows
that agreement. Because of the concurrence between the two measures of peak
intensity, it has to be concluded that the SPFA matched the experiment and the
SEFA did not. The fact that the analytical SEFA approximation agreed with TDSE
occurs because TDSE also uses the DA, making the experiment a demonstration of
the superior accuracy of the SPFA.%

The conclusion that the SPFA performed significantly better than the SEFA in
an experiment done with a laser wavelength of $1.51\mu m$, suggests that the
known failure of the DA at low frequencies is already in evidence at this wavelength.

\subsubsection{SPFA-SEFA coalescence at high frequencies}

\begin{figure}
[ptb]
\begin{center}
\includegraphics[
height=4.5576in,
width=4.6449in
]%
{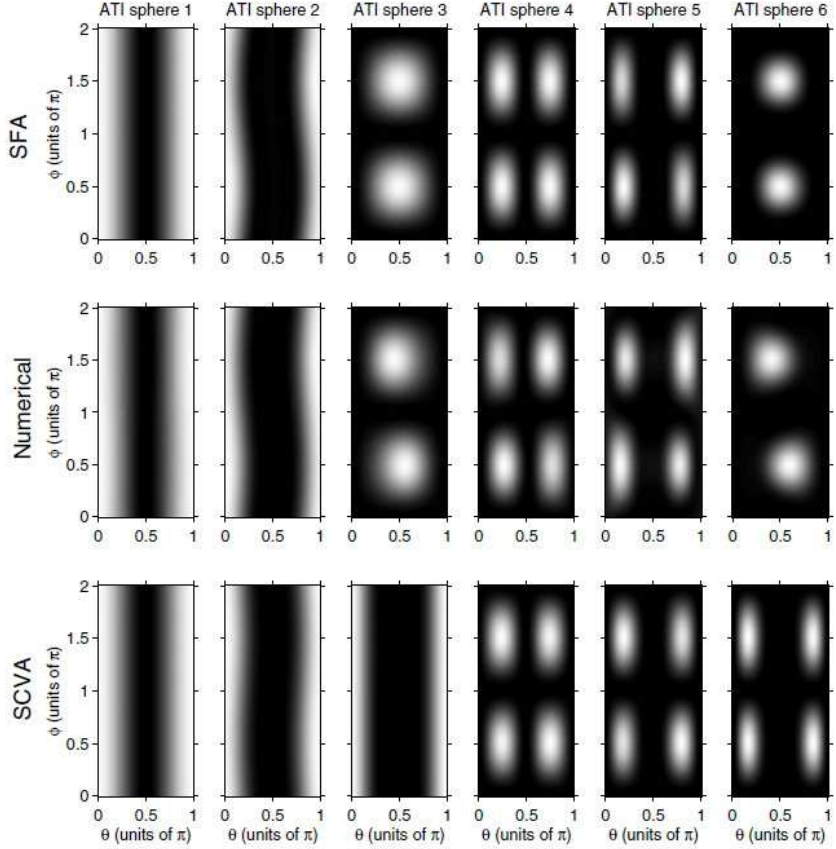}%
\caption{This is a reproduction of Fig. 1 in Ref. \cite{bondar}. A
superposition of two linearly polarized fields is analyzed, with $\omega
_{1}=1$ $a.u.$ and $\omega_{2}=3$ $a.u.$, and the intensity of each field is
$3.5\times10^{14}$ $W/cm^{2}$. For the significance of other labeling in the
figure, see Ref. \cite{bondar}. For present purposes, the important features
are the remarkably close correspondences between the panels labeled
\textquotedblleft SFA\textquotedblright\ (SPFA in this article) and those
labeled "Numerical\textquotedblright\ (TDSE).}%
\label{fig4}%
\end{center}
\end{figure}

An analytical demonstration that SPFA and SEFA methods coalesce at high
frequencies comes from the exact correspondence between the SPFA and the
Faisal approximation \cite{faisal} that is a high-frequency SEFA method. This
conclusion is borne out by high-frequency TDSE calculations reported in Ref.
\cite{bondar}. A sample of this agreement is shown in Fig. \ref{fig4}, with
other figures in Ref. \cite{bondar} showing equally good agreement. Further
verification of the high-frequency SPFA-SEFA coalescence was found
\cite{hrgav} in a comparison between the SPFA and the high-frequency method of
Gavrila \cite{pontgav,gavrev}.%

\subsubsection{Extension to relativistic conditions}

An important feature of the SPFA method of Ref. \cite{hr80} is that it is
evolved from a fully relativistic Klein-Gordon (i.e. spinless
\textquotedblleft electron\textquotedblright) formalism \cite{hr90}. That is,
there are actually no frequency limitations on the relativistically-extended
SPFA formalism, as long as the intensity is high. This conclusion is further
reinforced by fully Dirac-relativistic calculations for ground-state hydrogen
with circular polarization \cite{hrrel} and with linear polarization
\cite{dpcdiss}. These calculations have been shown to reduce exactly to the
nonrelativistic SPFA of Ref. \cite{hr80}. Dirac-relativistic SPFA calculations
can be done with any hydrogenic initial state, since analytical solutions of
the Dirac equation are known for hydrogen \cite{bethesal}. The algebraic
manipulations with Dirac matrices can be tedious, but the results are in
closed analytical form, meaning that numerical examples are simple to program.

Some practical features of the relativistic SPFA are shown in Ref.
\cite{hropex01}.

It has been emphasized above that the SPFA, since it relates directly to laser
fields rather than to a proxy for laser fields, provides valuable insights. An
important example is the analysis of the radiation pressure experiment in Ref.
\cite{smeenk}. With nonrelativistic strong-field conditions, a photoelectron
ionized by a circularly polarized laser will enter into a circular orbit
around the remnant ion, with the orbit lying in a plane perpendicular to the
propagation direction. This plane contains the ion as the origin of the
circular orbit. Radiation pressure will cause the plane of the circular orbit
to be projected slightly in the propagation direction. \ See Refs.
\cite{hropex01,hrsmeenk} for an illustration of this effect. A calculation of
the angle by which the plane of rotation is projected forward was attempted in
three publications \cite{smeenk,cbc,cbc2}, but without a successful outcome.
From the point of view of relativistic effects associated with laser fields,
the calculation is simple and insightful. Momentum associated with absorption
of circularly polarized photons is simply the sum of the spin momenta of the
photons since their momenta are aligned. Absorption of the photons necessary
to reach ionization threshold ($E_{B}+U_{p}$) is transmitted to the initial
atom, with negligible effect because of the atomic mass. The additional
photons required to achieve the energy $U_{p}$ of the photoelectron in its
circular orbit provide a momentum $U_{p}/c$ to the photoelectron in the
propagation direction. The plane of rotation is thereby tilted forward by an
angle%
\begin{equation}
\theta=\arctan\left(  p_{\bot}/p_{\Vert}\right)  =\arctan\left(  \frac
{U_{p}/c}{\sqrt{2mU_{p}}}\right)  =\arctan\left(  \sqrt{\frac{U_{p}}{2mc^{2}}%
}\right)  . \label{G}%
\end{equation}
This result is independent of the identity of the atom ionized, in agreement
with the intense-field part of the experiment, and is also quantitatively
correct, thus supplying confirmation of the predicted angle in Ref.
\cite{hrrel} [see Eq.(5.5)] that is identical to Eq. (\ref{G}). A result
unsuccessfully sought in three major publications by SEFA means, is found
correctly in one paragraph by SPFA means.

\section{Conclusions}

Strong-field laser interactions with matter have been analyzed primarily in
terms of the DA. This corresponds to replacement of the propagating laser
field by an oscillatory electric field. When the laser field is sufficiently
intense that it is the dominant influence on the photoelectron, the
relativistic character of the laser field, propagating at the speed of light
in vacuum, cannot be ignored. This infers the conclusion that a strong-field
approximation based in propagating fields (SPFA) is more efficient and
reliable than a strong-field approximation based on oscillatory electric
fields (SEFA).

Powerful evidence for this conclusion exists. A brief compendium follows.

\begin{itemize}
\item[$\checkmark$] First and foremost is that the ATI phenomenon, so
startling to a community accustomed to the DA, was well-understood and
predicted in advance of its laboratory observation when examined from a
propagating-field point of view. Quantitative advantages exist in addition to
qualitative factors. Some of the earliest precision laboratory results dating
from 1986 \cite{bucks} and 1993 \cite{moh} were successfully modeled by an
SPFA method \cite{hr80} that provides closed-form analytical results that are
simple to evaluate.

\item[$\checkmark$] The channel-closing phenomenon, where energy demands to
offset an increase in ponderomotive energy with intensity causes the minimum
photon order to index upwards, is often credited to Muller, Tip, and van der
Wiel in a 1983 paper \cite{mullertipvdw}. However, the earlier Ref.
\cite{hr80} contains an explicit treatment of this phenomenon. Apart from this
reference, channel closing was well known from the early work with the Volkov
solution \cite{hrdiss,hr62}, where the first channel closing is identified as
a sufficient condition for the failure of perturbation expansions. Thus, this
well-known phenomenon was reported as a basic consideration in strong-field
physics 25 years before it came to general attention in the laser-effects community.

\item[$\checkmark$] Adiabaticity, where decreasing laser frequency is regarded
as causing an approach to static conditions, was introduced into strong-field
physics by Keldysh in 1964. It was adopted by the strong-laser community from
the beginning of post-ATI activity in 1979, because it is a natural aspect of
the tunneling model. Adiabaticity was shown to be fallacious for propagating
wave phenomena in 2000 \cite{hr63}, 2008 \cite{hr101}, and 2014 \cite{hrtun}
articles, but this proof of failure has not stopped exploration of
adiabaticity by some researchers in strong-field physics \cite{santra}.

\item[$\checkmark$] A companion problem with adiabaticity is the requirement
that any acceptable theory of strong-field effects must have a static
zero-frequency limit. This contradicts the basic fact that laser fields always
propagate at the speed of light, no matter how low the frequency might be. The
demand for zero-frequency behavior is an especially pernicious fallacy in that
it labeled as unacceptable any SPFA theory. The zero-frequency criterion
remained in force continuously from 1979 until just very recently, and even
found its way into textbooks \cite{jkp} on strong-field laser effects. The
cost of rejecting any properly-posed theory for nearly 40 years is incalculable.

\item[$\checkmark$] The tunneling model, appropriate only for electric fields,
remains entrenched in the strong-field community. Articles, lecture notes, and
books on strong-laser phenomena routinely show a diagram of an attractive
Coulomb potential being tilted by the potential of an electric field. This is
so ubiquitous that laser-caused ionization is often regarded as being
identical with that illustration. The tunneling model has no meaning for
strong-laser ionization
\end{itemize}

The toll exacted by reliance on unphysical models is impossible to estimate,
but it has clearly caused needless expenditure of research resources. A major
example comes from continued focus on the tunneling model, even though the
tunneling model is strictly a DA concept that is inconsistent with
propagating-field behavior. Laser fields are vector fields that are not
consonant with tunneling behavior, and this vector property becomes
fundamental with intense fields. Nevertheless, exploration of tunneling
phenomena remains a major focus of funded research.

\end{document}